\documentclass[twocolumn]{aastex63}

\newcommand\kms{$\mathrm{km\,s}^{-1}\,$}
\received{June 1, 2019}
\revised{January 10, 2019}
\accepted{\today}
\submitjournal{AJ}

\shorttitle{Machine Learning Approach to Spectroscopic Observations}
\shortauthors{Rhea et al.}

\graphicspath{{./}{figures/}}
\usepackage{comment}
\begin{document}

\title{A Machine Learning Approach to Integral Field Unit Spectroscopy Observations: I. HII Region Kinematics}

\correspondingauthor{Carter Rhea}
\email{carterrhea@astro.umontreal.ca}

\author[0000-0003-2001-1076]{Carter Rhea}
\affiliation{D\'epartement de Physique, Universit\'e de Montr\'eal, Succ. Centre-Ville, Montr\'eal, Qu\'ebec, H3C 3J7, Canada}

\author[0000-0002-5136-6673]{Laurie Rousseau-Nepton}
\affiliation{Canada-France-Hawaii Telescope, Kamuela, HI, United
States}

\author[0000-0002-1755-4582]{Simon Prunet}
\affiliation{Canada-France-Hawaii Telescope, Kamuela, HI, United
States}

\author[0000-0001-7271-7340]{Julie Hlavacek-Larrondo}
\affiliation{D\'epartement de Physique, Universit\'e de Montr\'eal, Succ. Centre-Ville, Montr\'eal, Qu\'ebec, H3C 3J7, Canada}

\author[0000-0003-2239-7988]{S\'ebastien Fabbro}
\affiliation{NRC Herzberg Astronomy and Astrophysics, 5071 West Saanich Road, Victoria, BC, V9E 2E7, Canada}
\affiliation{Department of Physics and Astronomy, University of Victoria, Victoria, BC, V8P 5C2, Canada}

\begin{abstract}
 SITELLE is a novel integral field unit spectroscopy instrument that has an impressive spatial (11 by 11 arcmin), spectral coverage, and spectral resolution (R$\sim$1-20000). SIGNALS is anticipated to obtain deep observations (down to $3.6\times10^{-17}$ergs s$^{-1}$cm$^{-2}$) of 40 galaxies, each needing complex and substantial time to extract spectral information.
We present a method that uses Convolution Neural Networks (CNN) for estimating emission line parameters in optical spectra obtained with SITELLE  as part of the SIGNALS large program. 
Our algorithm is trained and tested on synthetic data representing typical emission spectra for HII regions based on Mexican Million Models database (3MdB) BOND simulations. 
The network's activation map demonstrates its ability to extract the dynamical (broadening and velocity) parameters from a set of 5 emission lines (e.g. H$\alpha$, N[II] doublet, and S[II] doublet) in the SN3 (651-685 nm) filter of SITELLE.
Once trained, the algorithm was tested on real SITELLE observations in the SIGNALS program of one of the South West fields of M33. 
The CNN recovers the dynamical parameters with an accuracy better than $5\, $\kms in regions with a signal-to-noise ratio greater than 15 over the H$\alpha$ line. 
More importantly, our CNN method reduces calculation time by over an order of magnitude on the spectral cube with native spatial resolution when compared with standard fitting procedures. 
These results clearly illustrate the power of machine learning algorithms for the use in future IFU-based missions.
Subsequent work will explore the applicability of the methodology to other spectral parameters such as the flux of key emission lines.
\end{abstract}


\section{Introduction} \label{sec:intro}
HII regions lay the foundation of many studies from star-formation in galaxies, to galactic evolution and cosmology, and are one of the main drivers of observational extra-galactic astronomy (e.g. \citealt{french_galaxies_1980}; \citealt{weedman_ngc_1981}; \citealt{veilleux_spectral_1987}). HII regions form when the gaseous clumps are irradiated by an interior young and hot star or cluster of stars causing the gas to become partially or completely ionized (e.g. \citealt{osterbrock_astrophysics_1989}; \citealt{shields_extragalactic_1990}; \citealt{franco_evolution_2000}). They are primarily composed of Hydrogen and Helium, but contain non-negligible amounts of metals and their ionized counterparts (e.g. \citealt{shields_composition_1976}; \citealt{oey_abundances_1993}; \citealt{kennicutt_systematic_1993}; \citealt{garnett_composition_1987}). The characteristic bright emission lines coming from recombination and collision between the free electrons and the different atoms/ions in the nebulae are observed at large distances and allow the study of interstellar matter and its primary constituents (e.g. \citealt{kewley_host_2006}; \citealt{crawford_rosat_1999}; \citealt{baldwin_classification_1981}). Additionally, the omnipresence of the HII regions in some galaxies allow for the study of galactic disk dynamics (e.g. \citealt{epinat_ghasp_2008}), magnetic fields and turbulence at large and small-scales (e.g. \citealt{odell_turbulent_1986}; \citealt{haverkorn_measuring_2015}; \citealt{beck_galactic_1996}; \citealt{quireza_electron_2006}; \citealt{pavel_h_2012}), and the importance of various feedback mechanisms that inject energy into the ISM, i.e. stellar winds, supernovae and radiation pressure (e.g. \citealt{mcleod_stellar_2020};
\citealt{ramachandran_stellar_2018,ramachandran_testing_2019}).
 
 More recently, the use of integral field spectroscopy on nearby galactic and extragalactic HII regions has offered a more complete view of their physical properties (e.g. \citealt{leroy_portrait_2016}; \citealt{sanchez_integral_2012}; \citealt{bundy_overview_2014}). Also, increasing spectral and spatial resolution has allowed for the study of the complex dynamical structures of the HII regions and pushed the limit of previous analysis methods meant for integrated/unresolved spectra of HII regions (e.g. \citealt{martins_near-ir_2010}; \citealt{sanchez_integral_2012}; \citealt{drissen_imaging_2014}). 
 Typical fitting procedures used to extract the dynamics and emission lines flux measurements from HII regions spectra require a good prior estimate of the velocity as well as the number of velocity components to be fitted (e.g. \citealt{zeidler_young_2019}; \citealt{bittner_gist_2019}; \citealt{sanchez_pipe3d_2007}). 
 Defining the range of those priors is usually not a problem when the ensemble of spectra shows similar characteristics. While the typical range of velocity seen in galactic disks can easily vary by a few hundreds of \kms (e.g. \citealt{dressler_rotational_1983}; \citealt{bregman_galactic_1980}; \citealt{sancisi_cold_2008}), and the internal dynamics of HII regions can add thermal/turbulent broadening and expansion velocity to the galactic contribution (e.g. \citealt{sofue_galactic_1995}; \citealt{arsenault_integrated_1986}), the typical velocity prior for a given spectral data cube can be very broad and is often not precise enough to ensure a proper fit of the entire data set. We are additionally facing new challenges in the dynamical analysis, because the spatially resolved HII regions spectra often contain emission from different phases of the ISM (along the line of sight) and can be composed of multiple dynamically distinct components (e.g. expanding shells, \citealt{rozas_h_2007}; \citealt{relano_expansive_2005}) having each a different thermal/turbulent broadening. Of course, fitting two or more components with the proper velocity and broadening priors is the best approach in such case, but only when such components are actually present in the spectra (e.g. \citealt{relano_internal_2005}; \citealt{le_coarer_e_halpha_1993}). 
 
 Ultimately, extracting the information in a consistent manner from high spectral and spatial resolution data cubes requires a dedicated method to estimate the priors on the different spectral parameters, taking into account the variation of the observed spectral features across the field-of-view. 

SITELLE, the Imaging Fourier Transform Spectrograph (IFTS) of the Canada-France-Hawaii Telescope (CHFT), produces spectral data cubes containing over 4 million pixels with adjustable resolving power (up to 10,000) and has an instrumental line shape described by a sine cardinal function (\citealt{martin_calibrations_2017}; \citealt{baril_commissioning_2016}; \citealt{drissen_sitelle_2019}). Its 11$^{\prime}$\,$\times$\,11$^{\prime}$ field-of-view (FOV) contains more than 4 million pixels for which the spectral sampling and resolution varies as a function of their relative position angle with the mobile mirror. Moreover, emission lines intensities (and therefore line intensity ratios) may vary significantly across the parameter space of the physical properties observed in HII regions. 

All together, these characteristics make a typical template fitting strategy (e.g. cross-correlation function maximization) very difficult to implement since the sine cardinal function side lobes affect neighbouring line intensity and shape, and the position of the lobes with respect to the central position of the line varies with spectral resolution (changes across the FOV). In addition, the variation of line intensity ratios between different emission regions can lead to gross errors on the velocity estimates when a single template spectrum is used. Therefore, an adapted approach is developed here to solve these issues while still fitting entire data cubes, using the same uniform and reproducible method and including the dynamical and spectral complex nature of the resolved HII regions. 

This paper explores the use of a Convolution Neural Network to resolve deficiencies in the existing fitting software \texttt{ORCS} -- \textit{Outils de R\'eduction de Cubes Spectraux}. Although the \texttt{ORCS} fitting routines are robust, they require a human-generated prior for all fits; this paper demonstrates the use of machine learning to estimate the priors with no human input. In $\S$ 2, we outline the Convolution Neural Network and the synthetic data set used to train the network. We explore the success of our CNN to the synthetic data in $\S$ 3.  In $\S$ 4, we discuss the applicability of our methodology to low resolution spectra. Additionally, we apply the CNN to a field of M33 in order to test its efficacy in real observations. Finally, in $\S$ 5, we recap the main successes and outline our future work. 

\section{Methodology} \label{sec:meth}
\subsection{Convolutional Neural Networks} \label{sec:cnn}
\begin{figure*}[th!]
    \centering
    \includegraphics[width=0.9\textwidth]{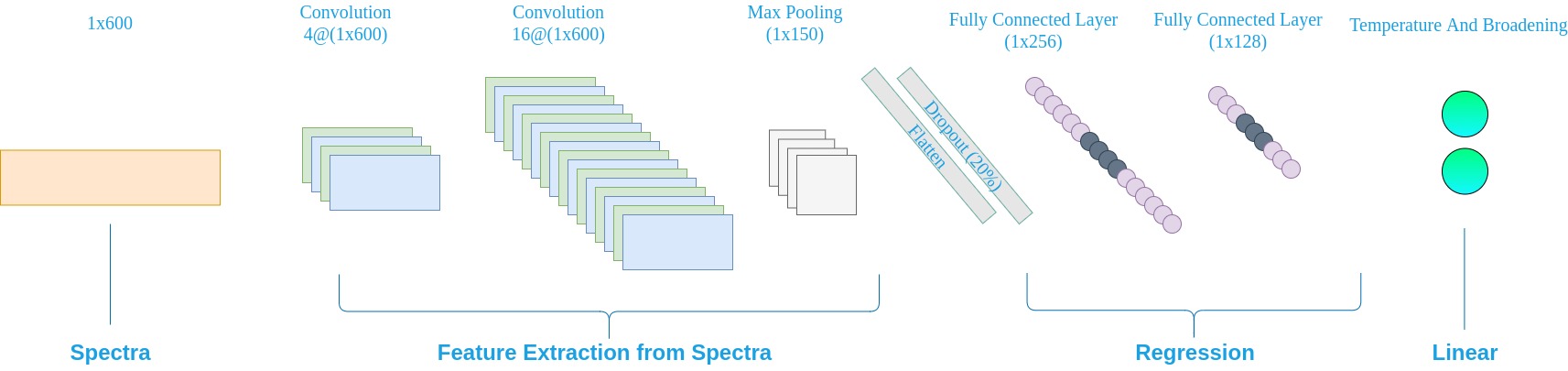}
    \caption{A cartoon of the convolutional neural network used in this work. As described in the text, it is an adaptation of the \texttt{STARNET} topology (\citealt{fabbro_application_2018}). The input spectra is first convolved in two separate layers before being condensed in a pooling layer. Once flattened, the vector is passed to two hidden layers. Finally, the velocity and broadening parameters are estimated using two separate output nodes denoted by the blue-green bar.}
    \label{fig:CNN}
\end{figure*}
Neural Networks have been used extensively in astronomy to classify galaxies (\citealt{storrie-lombardi_morphological_1992}), separate galaxies from stars (\citealt{bertin_classification_1994}), categorize dynamic parameters of galaxy clusters (e.g. \citealt{ntampaka_dynamical_2016}; \citealt{ntampaka_deep_2019}), explore astrophysical morphologies at differing scales (e.g. \citealt{sadaghiani_physical_2019}; \citealt{iwasaki_x-ray_2019}), derive galaxy redshift from wide band images (\citealt{pasquet_photometric_2019}), and extract emission-line parameters from spectra (e.g. \citealt{olney_apogee_2020}; \citealt{ucci_interstellar_2019}; \citealt{baron_machine_2019}). A recent effort to calculate the parameters of HII regions from their spectra, \texttt{GAME}\footnote{\url{https://game.sns.it/}}, employs a combination of Decision Trees and AdaBoost in order to predict physical parameters (\citealt{ucci_inferring_2017}; \citealt{ucci_game_2018}). 
In lieu of this, our method uses a Convolutional Neural Network (CNN) architecture designed by \cite{fabbro_application_2018}, monikered  \texttt{STARNET}, which has already demonstrated success in estimating emission-line parameters from stellar spectra.  

During the course of this work, we became aware of the work of \cite{keown_clover_2019}, which uses an approach similar to ours to estimate the velocity and broadening of high resolution radio emission lines, taking into account possible multiple velocity components. While their work focuses on high resolution, isolated emission lines, ours focuses on lower resolution spectra observed on a wide field of view, hence often with a wide velocity distribution. In addition, the \texttt{SITELLE} ILS extended structure prevents us in any case from considering the different emission lines separately.

Our convolutional neural network is graphically depicted in figure \ref{fig:CNN} and laid out as follows:
\begin{enumerate}
    \item 8x8 convolution with 4 filters
    \item 4x4 convolution with 8 filters
    \item Global max pooling with 4 filters
    \item 20$\%$ dropout 
    \item 256 fully-connected nodes
    \item 128 fully-connected nodes
    \item 2 output neurons
\end{enumerate}

The CNN takes the normalized SITELLE emission spectra obtained with the SN3 filter (651-685 nm) and returns an estimate on the velocity (\kms) of the lines and their broadening (\kms), assuming they are consistent over the five major emission lines in SN3.  We tested several scaling functions (RobustScaler, StandardScaler, and MinMaxScaler); although we obtained the tightest constraints with the MinMaxScaler, the activation map revealed fitting nonphysical features and noise. We therefore normalize the spectrum to have a maximum value equal to unity.

In order to ensure the appropriate hyper-parameters, we explored their spaces extensively using the random search algorithm, as implemented by \texttt{sklearn}, embedded in a 10-fold cross correlataion. Throughout our training, we saw no significant deviation from the results reported by \cite{fabbro_application_2018}. Therefore, we adopted the same hyper-parameter values as used in the standard \texttt{STARNET} procedure. Structural hyper-parameters can be readily seen in figure \ref{fig:CNN}. In order to view the other parameters (i.e. learning rates, decay rates, etc.), we suggest the reader view our github page: \url{https://github.com/sitelle-signals/Pomplemousse}. We report a maximum number of 10 epochs and an initial batch size of 8 spectra.

\subsection{Synthetic Data}\label{sec:syn}
In order to demonstrate the feasibility of using a CNN to identify the correct spectral parameters, we construct a set of synthetic data on which to train and test the network.
The synthetic data set used in this study was created using the \texttt{ORB} software developed to reduce data from \texttt{SITELLE} (\citealt{martin_optimal_2016}). To generate synthetic spectra, We use the \texttt{ORB create\_cm1\_lines\_model} function which requires a number of parameters that will be defined in this section.
Since our tool was developed primarily for SITELLE's programs and the SIGNALS collaboration, we focused on the SN3-filter which covers a band pass between 647 and 685 nm. In accordance with the SIGNALS survey, we select a primary spectral resolving power of 5000, an exposure time of 13.3s per step, and 842 steps (\citealt{rousseau-nepton_signals_2019}). In order to replicate the change of spectral resolution across the cube, we allow the resolving power to randomly vary between 4800 and 5000 since the resolution will vary between these values in any given SN3 observation which is a part of the SIGNALS program. We will model the following lines: [NII]$\lambda6548$, H$\alpha(6563)$\AA, [NII]$\lambda6583$, [SII]$\lambda6716$, and S[II]$\lambda6731$. Furthermore, we use the \texttt{sincgauss} function as described in \cite{martin_optimal_2016} to include line broadening. We randomly varied the velocity between -200 and 200 \kms, while the broadening was randomly varied between 0 and 50 \kms. These ranges were selected from our prior knowledge of the distribution of velocities in M33 (\citealt{epinat_ghasp_2008}) and the typical broadening in SITELLE data cubes at this spatial resolution. Note that we randomly selected the resolution, broadening, and velocity parameters with replacement for each synthetic spectrum.
The final input required to construct the synthetic spectra is the amplitude of each emission line.

In order to calculate reasonable relative fluxes for the five lines while ensuring we are sampling the desired physical parameter space, we used the 3MdB\footnote{\url{https://sites.google.com/site/mexicanmillionmodels/}} -- Mexican Million Models Database (\citealt{morisset_virtual_2015}). The 3Mdb contains models created using the \texttt{CLOUDY v17.01} photoionization code based on a pre-selected set of emission region parameters and underlying ioinizing stellar spectra (\citealt{ferland_2017_2017}).
We use the \texttt{BOND} dataset described in \cite{asari_bond_2016} which contains spectra from HII regions similar to those expected to be found in SIGNALS. The \texttt{BOND} data-set contains 63000 spectra. Though the data set covers the physical parameter space of the emission nebulae we wish to study, it also contains a number of models that are outside the scope of our study. 
We describe varying parameters used in table \ref{tab:CloudyParameters}. While the \texttt{BOND} simulations have two simulation geometries, completely filled and thin shell, we remove all thin shell (fraction=0.03) simulations from our sample. This leaves us with filled spheres with a density of approximately 100 cm$^3$ and represents a younger population of HII regions (e.g. \citealt{asari_bond_2016}; \citealt{stasinska_excitation_2015}; \citealt{cedres_filling_2013}). 

\begin{table}[!h]
    \centering
    \begin{tabular}{|c|c|c|c|}
        \hline 
        Parameter & Lower Limit & Upper Limit & Step Size \\
        \hline \hline 
        log(U) & -3.5 & -2.5 & 0.5 \\ \hline
        Age (Myr) & 1 & 6 & 1 \\ \hline
        12+log(O/H) & 7.4 & 9.0 & 0.2 \\ \hline
        log(N/O) & -2 & 0 & 0.5 \\ \hline
    \end{tabular}
    \caption{HII region parameter selection used during the M3db runs of the \texttt{BOND} simulations. The initial run-parameters were cut further in order to focus on the emission expected in the SIGNALS program. The step sizes were set by the 3Mdb runs (see \cite{morisset_virtual_2015} for more information)}
    \label{tab:CloudyParameters}
\end{table}

\begin{figure}[h!]
    \centering
    \includegraphics[width=0.45\textwidth]{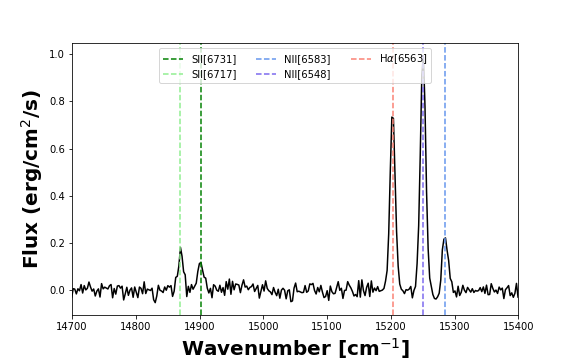}
\caption{Example spectrum simulated using the process described in $\S$\ref{sec:syn}. As our population statistics suggest, this is not the only expected spectral shape. However, it is representative of the sample and clearly demonstrates the five emission line peaks. This is the SN3 spectral coverage of SITELLE.}
    \label{fig:ex_spec}
\end{figure}

We further constrained the ionization parameter, U, and metallicity proxy, 12+log(O/H), to focus on SIGNALS-type HII regions (e.g. \citealt{rousseau-nepton_signals_2019}; \citealt{perez-montero_revisiting_2019}; \citealt{kashino_disentangling_2019}; \citealt{zinchenko_effective_2019}). 
With these constraints, we extracted the amplitudes of the five emission lines present in SN3, first randomly selecting a model which passed our selection criteria. We then normalized the amplitudes with respect to H$\alpha$. After combining the five lines (with the appropriate instrumental line shape) and the simulated continuum emission, we add a noise component. The SNR is sampled from a uniform distribution between 5 and 30. Below a SNR of 5, the lines are nearly indistinguishable and the sidelobes of the ILS are completely obstructed. We expect a nominal high ($>20$) SNR for H$\alpha$ in the SIGNALS program. 
SNR effects will be investigated later in the article. Figure \ref{fig:ex_spec} shows a sample spectrum. At this stage, we create 50,000 mock spectra in the form of \texttt{FITS} files which contain the emission parameter information (e.g. velocity, broadening, resolution).

\begin{figure*}
    \plottwo{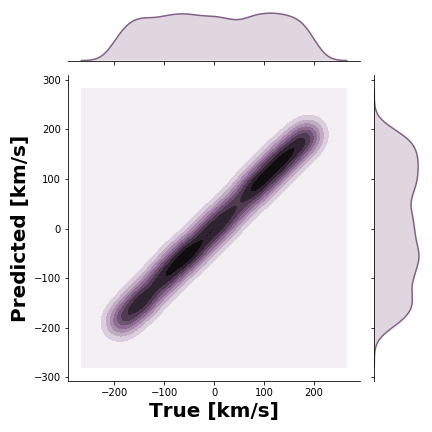}{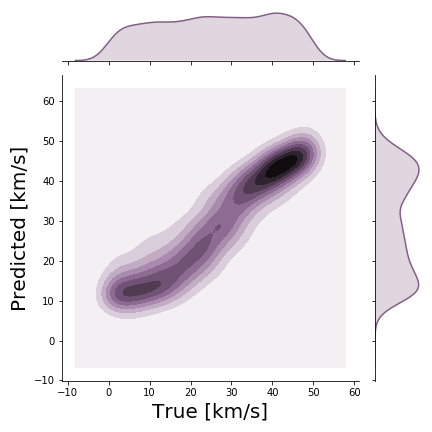}
    \caption{Kernel Density Estimation (KDE) plots for the test set. Left: True vs Predicted Velocity values in \kms. Right: True vs Predicted Broadening values in \kms. In both plots we can see that the predicted values accurately mimic the true values. Note the change in scales between the two plots.}
    \label{fig:KDE_plots}
\end{figure*}

\begin{figure*}
    \plottwo{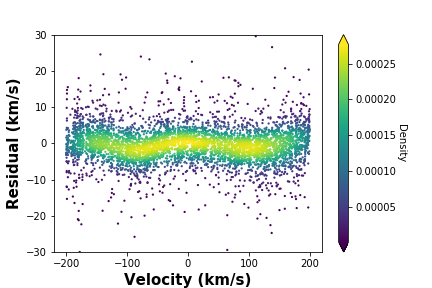}{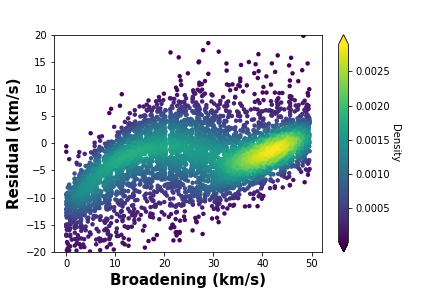}
    \caption{Left: Velocity Residual as a function of the true velocity. Although there exists a background substructure, it only affects a fraction of a percent of the total test set and is thus negligible. Right:Broadening residual as a function of the true broadening. The pattern demonstrates a bias for low broadening values that is likely caused by the networks inability to distinguish a low amount of broadening. Moreover, the broadening naturally segregates itself into two physical peaks typical of HII regions and supernovae remnants, respectively (e.g. \citealt{veilleux_spectral_1987}; \citealt{vasiliev_velocity_2015}).
    }
    \label{fig:Resid_plots}
\end{figure*}

\begin{figure*}
    \plottwo{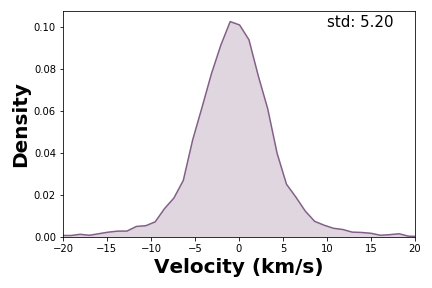}{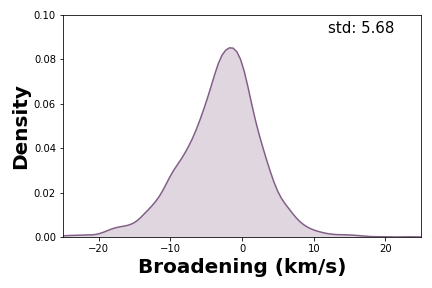}
    \caption{Left: Density plot of the velocity residuals in \kms along with the standard deviation. Right: Density plot of the broadening residuals in \kms in addition to the standard deviation. The asymmetry is likely due to the diversity of resolving power introduced in the training set.
    }
    \label{fig:Dist_plots}
\end{figure*}


\subsection{SITELLE Data}
\subsubsection{Calibration and Data Reduction}
Observations of M33 were taken during the Queued Service Observing period 18B (Program 18BP41, P.I. Laurie Rousseau-Nepton) at the Canada France Hawaii Telescope on the summit of Mauna Kea, Hawaii, using SITELLE. These exposures were taken with the SN3 filter which covers a range from 651-685 nm for a total of 4h with a spectral resolving power of R$\sim$5000. The pointing was centered on a single field in M33 and is part of a larger observation of M33 in its entirety. This observation also forms a basis for the SIGNALS program, lead by Laurie Rousseau-Nepton, which aims to further categorize HII and star-forming regions in nearby galaxies. We note that the authors of this paper are members of the SIGNALS collaboration. 

The raw data were reduced and calibrated using SITELLE's personalized software, ORBS (version 3.1.2 \citealt{martin_optimal_2016}). We are able to resolve five spectral emission lines from our observations: [SII]$\lambda$6713, [SII]$\lambda$6731, [NII]$\lambda$6548, H$\alpha$, [NII]$\lambda$6584. Using the function \texttt{SpectralCube.Map\_Sky\_Velocity()}, we fit the OH sky line velocities, assumed at rest w.r.t. the observer, with a geometric model of the interferometer; afterwards, we used the function \texttt{SpectralCube.Correct\_Wavelength()} to refine the wavelength calibration of our data cube using the OH-lines fit.

\begin{figure}[h!]
    \centering
    \includegraphics[width=0.48\textwidth]{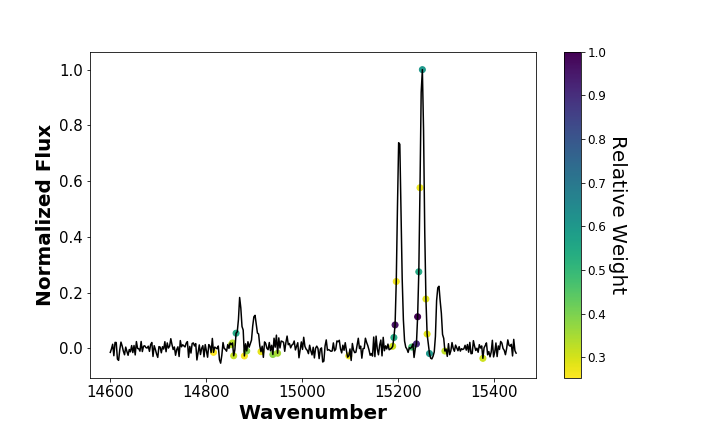}
    \caption{Activation or Saliency Map of our convolutional neural network applied to an example spectrum. The colored points represent the exact locations of the nodes in the input spectrum. Their color indicates their relative weight in the network. Weights under 0.25 are not shown for clarity.}
    \label{fig:activation}
\end{figure}

\section{Results} \label{sec:res}
In this section we apply our convolutional neural network outlined in $\S$\ref{sec:cnn} to our synthetic spectra with a resolution R$\sim$5000. We retained 70\% (35,000) of the spectra as our training set, 20\% (10,000) as our validation set, and the remaining 10\% (5,000) as the test set (e.g. \citealt{tetko_efficient_1997}). Training and validating our algorithm results in over 95\% accuracy for both predicted parameters: the velocity and the broadening. Accuracy is defined as the ratio of correct parameter estimations to the total number of estimates. An estimate is considered correct if it agrees with the ground truth value up to two digits after the decimal (i.e. to the hundredth place). The combined mean absolute error, another common metric for regression tasks, is $~5$kms$^{-1}$.  
Figures \ref{fig:KDE_plots}, \ref{fig:Resid_plots}, and \ref{fig:Dist_plots} visually depict the accuracy of the CNN on the test set and the associated residuals, respectively. As the figures depict, the algorithm was well trained and is able to accurately predict both the velocity and the spectral broadening. As evidenced in figures \ref{fig:KDE_plots} and \ref{fig:Resid_plots}, the predicted values are close to the ground truth values. The KDE plots in figure \ref{fig:KDE_plots} demonstrate that the parameter space is being well sampled for both the velocity and broadening. Figure \ref{fig:Dist_plots} demonstrates the Gaussian distribution of errors about zero; although the right panel reveals the slightly skewed error distribution of the broadening parameter, the shape is globally Gaussian and any distortion is believed to be caused by asymmetries within the training set.  We report a standard deviation of $\sim 5$ \kms for the velocity parameter.

This is well within the required limits as described in \cite{martin_optimal_2016} and \cite{rousseau-nepton_signals_2019} for an initial guess to be supplied to the \texttt{ORCS} software. The velocity error is required to be less than the channel width with corresponds to approximately 40 \kms for a resolution of 5000. The standard deviation of the broadening parameter is $\sim 5.5$ \kms. Since SITELLE resolves the broadening parameter down to approximately 3 \kms for high SNR regions ($\sim$1000), our broadening errors are near SITELLE's resolving power.

In order to compare the network results to those recovered by the \texttt{ORB}/\texttt{ORCS} software, we fit the test set using the \texttt{fit\_lines\_in\_spectrum} routine. The velocity and broadening parameters were initialized as the precise velocity and broadening parameters used to construct the spectra. Although this is improbable to occur during a standard fitting procedure, hence the need for an accurate estimate, this demonstrates the best possible case for the fitting algorithm. All other parameters were also set to those used to simulate the spectra. The fitting procedure recovers the true velocity with a standard deviation of $\sim3$\kms and the broadening with a standard deviation of $\sim4$\kms. Comparing these standard deviations with those from the CNN, we note that the \texttt{ORB}/\texttt{ORCS} recover the true parameters with marginally better accuracy.

Although the spread of errors shown in the figures \ref{fig:Dist_plots} and \ref{fig:Resid_plots} do not reveal overt overfitting, we applied a standard k-fold cross-validation algorithm  on ten partitions of the training, validation, and test data (e.g. \citealt{picard_cross-validation_1984}; \citealt{bengio_no_2004}). 
Overfitting occurs when the neural network learns the training set too well and is unable to generalize to other data sets such as the test set. Overfitting would manifest itself in these figures if they demonstrated a large spread of residuals (i.e. large errors).
We also implemented a modified k-fold cross-validation algorithm in which we varied only the training and validation data while retaining the same test set. We report approximately the same accuracy values (within 5\%) regardless of the fold and cross-validation technique. This further indicates the absence of overfitting (e.g. \citealt{cawley_over-tting_2010}; \citealt{molinaro_prediction_2005}). 

Additionally, we created an saliency map of our example spectrum from figure \ref{fig:ex_spec} which can be seen with the filled circles in figure \ref{fig:activation}. The saliency map delineates the regions of the input (in this case the spectrum) used by the convolutional neural network to learn (e.g. \citealt{simonyan_deep_2014}) by calculating the gradient of the output with respect to the input. 
More precisely, the map is created by varying one input variable at a time and calculating the change in the loss function. In this manner the algorithm highlights the most important input nodes.
We can clearly see by the clustering of data points in the image around the H$\alpha$ and [NII]$\lambda6548$ lines that the network considers these lines to be the most important components for determining the velocity and broadening. This is consistent with our expectations since these two lines, unlike the others, are consistently above the continuum in HII regions. It is sensible that the network does not weigh the [SII] doublet heavily since they are often unobservable due to noise. Moreover, the network does not focus only on the peaks of the H$\alpha$ and [NII]$\lambda6548$ lines, but also on their base. This indicates that the widening of the lines -- which is directly affected by the velocity and broadening components -- plays a crucial role in parameter estimation, as expected.

\section{Discussion}
While in Section 3, we demonstrated that the CNN algorithm is capable of extracting the correct spectral parameters (velocity and broadening) of the H$\alpha$, N[II], S[II] lines for synthetic SITELLE observations, in this Section, we examine the versatility of the model and its robustness when applied to real SITELLE observations. We also discuss the novelty of using such CNN algorithms for IFU observations in general (i.e. from other telescopes, especially in context of upcoming 30 and 40 -m class telescopes.
 
\subsection{Versatility of the Model}
While this technique is developed for the SIGNALS collaboration science case, aiming to obtain IFU observations of dozens of nearby galaxies, and thus R$\sim$5000, we demonstrate its applicability to other studies of HII regions using SITELLE at various spectral resolutions. Since there exists a number of other SN3 observations which are not a part of the SIGNALS program that were taken with an average spectral resolving power near R$\sim$2000, we wished to directly test our existing network and weights against synthetic data created with R$\sim$2000 (e.g. \citealt{puertas_searching_2019}; \citealt{gendron-marsolais_revealing_2018}; \citealt{rousseau-nepton_ngc628_2018}). However, since the resolution sets the number of steps (i.e. data points) in our spectrum, a reduction of the resolution affects the length of the input data. In order to feed lower resolution spectra into our CNN, we would be required to smooth or interpolate the data so that we would have an input of an equivalent length -- a requisite for use in a CNN. In doing so, we would be assuming a form of the interpolation (i.e. linear, a higher-order polynomial, spline, etc.) which might inject non-physical and potentially biased information into the spectra (\citealt{horowitz_efffects_1974}; \citealt{scargle_studies_1982}; \citealt{schulz_spectrum_1997}). We therefore do not modify the spectra, but instead we create an entirely new set of training, validation, and test data using the same routines employed to create our high spectral resolution synthetic dataset with a resolution set to R$\sim$2000.

After creating 30,000 synthetic spectra with a lower spectral-resolution, we divided the set into the training (70\%), validation (20\%), and test (10\%) sets. After training and validating our convolutional neural network, we applied it on our test data. We report a nominal accuracy of both predictors (velocity and broadening) of 92\% compared to 95\% in the case of R$\sim$5000. The standard deviation of the errors for the velocity and broadening are 75 and 12 \kms, respectively. We ran both k-fold cross-validation algorithms and again found consistency across the accuracy predictors. The results are coherent with our supposition that the method would extend well to relatively low resolution spectra since, even at R$\sim$2000, we are able to reasonably resolve the emission lines. The reduced accuracy is reasonable since the emission lines are less well-resolved. 

We attempted to use the network to predict low resolution SITELLE spectra (R$\sim$1000); however, at this resolution, the lines are often indistinguishable and the algorithm fails to achieve high-fidelity results. Typical SITELLE's observing strategy for targets in the local Universe and for the SIGNALS project, have an increased spectral resolution for the H$\alpha$ filter (SN3) and often a lower resolution for other filters (typically  R$\sim$1000). The dynamical priors (velocity and broadening) can then be estimated using the higher resolution SN3 filter and applied on the other observations of the same field with the other filters.
Overall, our results demonstrate that a CNN network is capable of reliably estimating spectral parameters (velocity and broadening) in SITELLE synthetic observations at high (R=5000) and low (R=2000) resolution, but that beyond R = 1000-1500, it fails because of the poor quality of observations. In other words, these results not only demonstrate that machine learning algorithms can be used to estimate kinematic parameters, but they also demonstrate the techniques limitations.

\subsection{Validation on a real data-set: the case of M33}\label{sec:M33}
\begin{figure}[h!]
    \centering
    \includegraphics[width=0.45\textwidth]{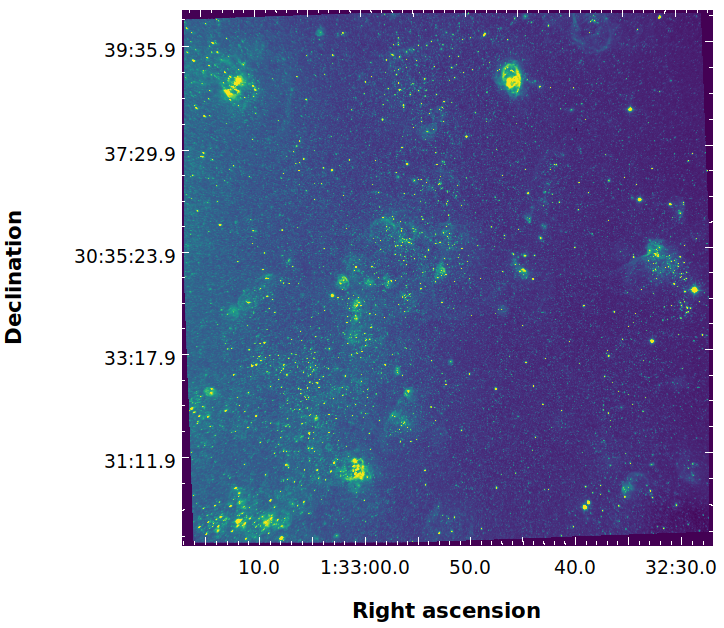}
    \caption{Deep, co-added SITELLE observation (4hr) of M33 Field 7 using the SN3 filter. The image illustrates the density of emission-line regions in the outskirts of M33.}
    \label{fig:DeepM33}
\end{figure}

With the ability of the CNN to predict velocity and broadening parameters accurately for synthetic data, we apply our methodology to an emission region of M33's South-East field (figure \ref{fig:DeepM33}). This region is an excellent test-bed for our algorithm since it contains several types of emission regions (i.e. HII region, planetary nebulae, etc.) and is part of the SIGNALS survey.

\begin{figure*}
    \plottwo{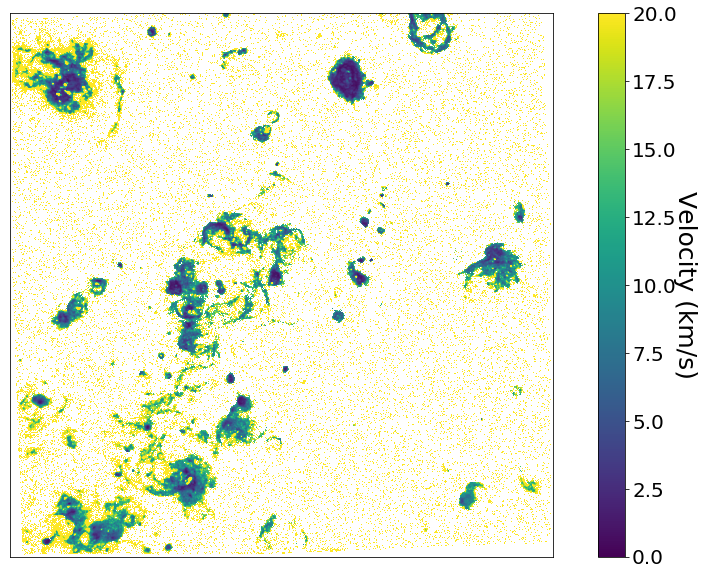}{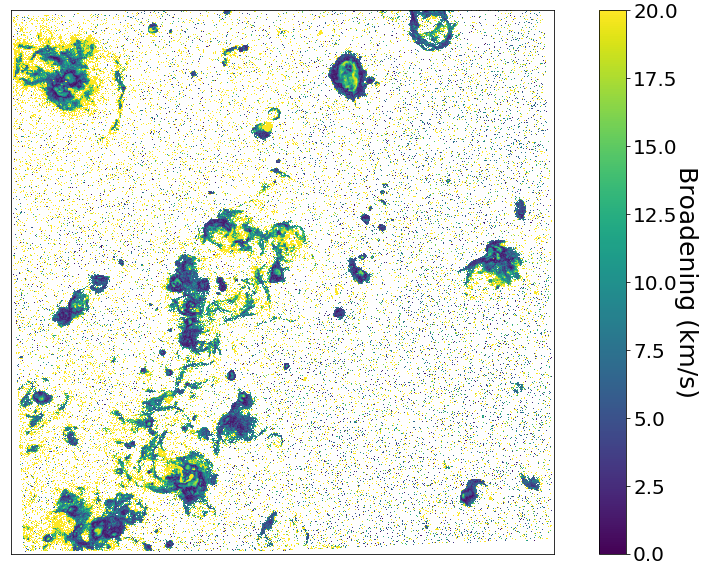}
    \caption{Left: Residual map of the velocity calculated from the absolute difference between the final \texttt{ORCS} fit and the machine learning priors calculated on an unbinned cube. Right: Residual map of the broadening calculated from the absolute difference between the final \texttt{ORCS} fit and the machine learning priors calculated on an unbinned cube. Both maps were smoothed using a 2-dimensional Gaussian kernel with a sigma value equal to 2 pixels.
    }
    \label{fig:Residuals}
\end{figure*}

Fits were calculated using the  \texttt{ORCS fit\_lines\_in\_region()} command centered on our five lines. 
Each grouping ([SII]$\lambda$6713/[SII]$\lambda$6731, [NII]$\lambda$6548/[NII]$\lambda$6584, and H$\alpha$) was fit simultaneously with a Gaussian convolved with a sinc function following the standard SITELLE procedure (\citealt{martin_calibrations_2017}); All lines were tied together with respect to the velocity and broadening.
Fits were optimized using the Levenberg-Marquardt least-squares minimization algorithm. In order to execute a fit in \texttt{ORCS}, the user is required to input an initial guess for the velocity and broadening parameters; this is due to the nature of the minimization algorithm.
The first set of priors were created by initially binning our cube into spatial bins of 8x8 followed by the standard \texttt{ORCS} fitting procedure. This standard method still requires an initial guess that the user must input. However, the machine learning method for determining priors does not require any user input and can be applied directly on the unbinned data. All fits were run using a computing server located at the CFHT headquarters in Waimea, Hawaii named \textit{iolani}. The server has 2 Intel \texttt{XEON E5-2630 v3} CPUs operating at 2.40GHz with 8 cores each. The configuration also has 64 GB of RAM available for computing purposes. 

A key benefit of the machine learning prior fits over the standard procedure is the economy of time associated with the machine learning algorithm. 
Since no fitting and iterating is necessary, the calculation time scales approximately linearly with the number of spectra.
Using a coarse initial binning, 8x8, the standard algorithm to calculate the priors takes approximately 4 hours in order to cover the entire cube. However, the unparallelized machine learning algorithm takes only 180 seconds\footnote{assuming a near-perfect speedup, we expect the parallelized algorithm to take approximately 25 seconds to run on \textit{iolani}.} to cover the same binned cube. Hence the machine learning algorithm calculates the priors more than 100 times faster than the standard algorithm.
We also calculate the time the machine learning algorithm takes to estimate the velocity and broadening parameters for an unbinned cube; this takes approximately 4 hours -- the same amount of time to calculate the standard priors on an binned (8x8) cube.

In addition to being considerably faster when estimating the priors, the machine learning algorithm also obtains accurate estimates. In order to quantify this notion, we calculate the residual values over the cube between the unbinned final fits -- using an 8x8 machine learning prior -- and the unbinned machine learning estimates. 
We only retained pixels for the residual analysis which demonstrated a flux value above our threshold of $2\times10^{-17}$ ergs/s. This threshold was chosen since it masks out all \texttt{nan} values and maintains the regions with clear emission. Figure \ref{fig:Residuals} demonstrates that the residuals are low in central parts of the emission regions, where the signal-to-noise is high, while the residuals are higher in the outskirts where the signal-to-noise is low. This is likely due to the fact that our synthetic data was created using a high signal-to-noise ratio of 50; we will explore the effects of the SNR ratio in a future paper.
While it is often desirable to study the emission in the outskirts in addition to the central emission, the low-residual regions outline locations of high-fidelity fits. In order to recover the velocity and broadening parameters in these regions, the machine learning estimates on either the binned or unbinned cube can be used as priors for a standard \texttt{ORCS} fit. Moreover, since the standard prior calculation requires binning spatially, substructure information is inherently lost in these priors. On the other hand, the convolutional neural network priors do not require any binning and thus retain all structural spatial information. 

\begin{figure}[h!]
    \centering
    \includegraphics[width=0.45\textwidth]{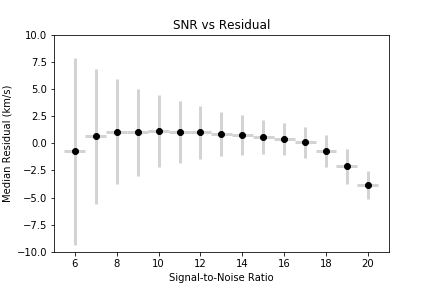}
\caption{Proxy signal-to-noise ratio versus mean absolute velocity residual (\kms) for the South West field of M33. For each SNR bin, we excluded outliers before calculating the mean absolute residual and standard deviation (grey y-axis error bars). Each SNR bin has a width of 1. 
}
    \label{fig:snr}
\end{figure}

\begin{figure}[h!]
    \centering
    \includegraphics[width=0.45\textwidth]{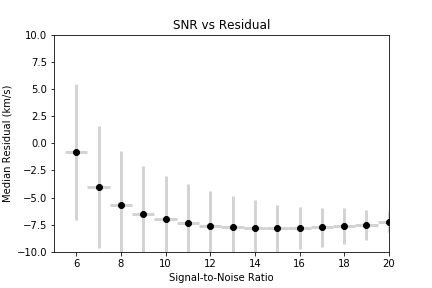}
\caption{Proxy signal-to-noise ratio versus mean absolute broadening residual (\kms) for the South West field of M33. For each SNR bin, we excluded outliers before calculating the mean absolute residual and standard deviation (grey y-axis error bars). Each SNR bin has a width of 1. 
}
    \label{fig:snr_broad}
\end{figure}

Although we do not study all the complexities of the SNR impact on our CNN in this article, we include a short discussion on it here. We calculate the SNR by dividing the H$\alpha$ flux by its fit uncertainty as calculated in our final \texttt{ORCS} fit. Although this is not exactly the SNR, it acts as a proxy value.
With the residual maps and the SNR proxy map, we have the residual and signal-to-noise information for each pixel. We then binned residuals by signal-to-noise ratio with a step size of 1 between 5 and 20. Twenty is the maximum value of the SNR proxy and below 5 we do not see any coherent structure in the spectra. We culled outliers that were outside of the 3-$\sigma$ range. Finally, we calculated the median absolute residual and standard deviation in each SNR bin.
As evidenced by figure \ref{fig:snr}, the accuracy of the CNN increases as the signal-to-noise ratio rises, an expected trend. 
Figure \ref{fig:snr_broad} demonstrates that the broadening residual plateaus at a SNR of approximately 12; moreover, the figure indicates a discordance between the CNN's estimations and those obtained from \texttt{ORCS} fits. We believe this behavior is due to the presence of multiple emission components serendipitously located in high SNR regions (see appendix for discussion). Multiple components affect the broadening parameter stronger than the velocity estimates. Even in standard fitting procedures, this poses a serious issue.

\subsection{Universal Applicability}

The methodology described in this paper is not limited to SITELLE data cubes. Indeed, the methodology naturally lends itself to any IFU-like data cube in which the observer has access to high-resolution spectral data such as the K-band Multi Object Spectrograph, KMOS (e.g. \citealt{sharples_first_2013}), or the Multi Unit Spectroscopic Explorer, MUSE (e.g. \citealt{bacon_muse_2010}). Since the machine learning algorithm is able to achieve reasonable estimations of the kinetic parameters (velocity and broadening) in a fraction of the time the standard fitting procedures take, it will play a crucial role in upcoming missions aimed at completing large-scale surveys using IFUs such as the Near-Infrared Spectrograph, NIRSpec (e.g. \citealt{de_oliveira_preparing_2018}),  on the James Webb Space Telescope and the MEGARA -- Multi-Espectrógrafo en GTC de Alta Resolución para Astronomía -- instrument on the Gran Telescopio Canarias (e.g. \citealt{paz_megara_2012}).

\section{Conclusions}

A convolution neural network has been exploited in several astronomical applications ranging from dynamic mass estimates of galaxy clusters (e.g. \citealt{ntampaka_deep_2019}) to the extraction of spectral parameters (e.g. \citealt{fabbro_application_2018}). This work applies a modified \texttt{STARNET} architecture (\citealt{fabbro_application_2018}) to high resolution (R$>$2000) SITELLE observations of HII regions in order to estimate the velocity and broadening parameters. Training, validation, and testing the machine learning algorithm with synthetic data integrating the 3Mdb database (\citealt{morisset_virtual_2015}) demonstrates the feasibility of the method. We demonstrate that the algorithm fails to predict the spectral parameters for low resolution (R$\simeq$1000) observations. We believe this is due to the lack of resolved spectral information resulting in partial blending of the main emission lines. However, above R$\sim$2000, we are able to disentangle the lines better. We apply the convolutional neural network to the Southwest field of M33 to calculate the velocity and broadening priors. Compared to the standard method for computing the priors, our method is over 100 times faster. Additionally, the machine learning algorithm can reliably estimate the emission-line parameters for the entire unbinned cube in roughly the same amount of time it takes the standard algorithm to calculate the priors on an 8x8 binned cube. 

The work presented here represents the first in a series of articles on the applications of machine learning to SITELLE spectra. In a subsequent article, we will present our work on the effects of the signal-to-noise ratio on convolution neural networks and how to mitigate the negative impacts.

We will also demonstrate the applicability of our methodology to calculate the fluxes (and ratios thereof) of emission lines, which will allow for the rapid classification of emission regions through grids of photo-ionization models (e.g. 3MdB). In the third proposed paper of the series, we will describe a machine learning methodology to identify possible multiple, blended components within emission lines.

\acknowledgments
The authors would like to thank the Canada-France-Hawaii Telescope (CFHT) which is operated by the National Research Council (NRC) of Canada, the Institut National des Sciences de l'Univers of the Centre National de la Recherche Scientifique (CNRS) of France, and the University of Hawaii. The observations at the CFHT were performed with care and respect from the summit of Maunakea which is a significant cultural and historic site.
C. R. acknowledges financial support from the physics department of the Universit\'e de Montr\'eal.
J. H.-L. acknowledges support from NSERC via the Discovery grant program, as well as the Canada Research Chair program.

The progamming aspects of this paper were completed thanks to the following packages of the python programming language (\citealt{van_rossum_python_2009}): \texttt{numpy} (\citealt{van_der_walt_numpy_2011}, \texttt{scipy} (\citealt{virtanen_scipy_2020}), \texttt{matplotlib} (\citealt{hunter_matplotlib_2007}), \texttt{pandas} (\citealt{mckinney_data_2010}), \texttt{seaborn} (\citealt{michael_waskom_mwaskomseaborn_2017}), \texttt{astropy} (\citealt{the_astropy_collaboration_astropy_2018}; \citealt{robitaille_astropy_2013}), \texttt{tensorflow} (\citealt{abadi_tensorflow_2015}, and \texttt{keras} (\citealt{chollet_keras_2015}).

\appendix
\section{SNR and the Residual}
As noted in $\S$\ref{sec:M33}, the broadening parameter (and the velocity parameter to a much lesser extent) exhibits an unexpected trend in its SNR vs residual plot (figure \ref{fig:snr_broad}). In this section, we explore potential reasons for this behavior: a dependence on the SNR of the training set, or an effect from multiple line components in high SNR regions. In order to determine whether or not the SNR of the training set has a negative impact on high SNR regions, we create a set of 1,000 synthetic data following the same prescription described before ($\S$\ref{sec:meth}); however, we allow the SNR to vary between 20 and 80 instead of stopping at 30. Because we are only created 1,000 synthetic spectra, we reduce the sampling rate of the velocity and broadening. This is not expected to have any effect on the results. We then apply our already trained network on the synthetic data. 
\begin{figure}[hb]
    \centering
    \plottwo{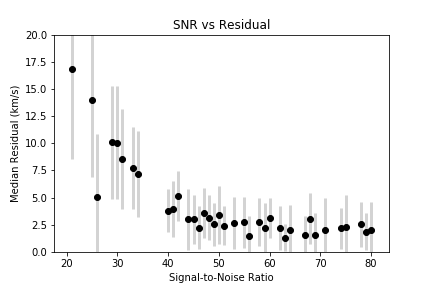}{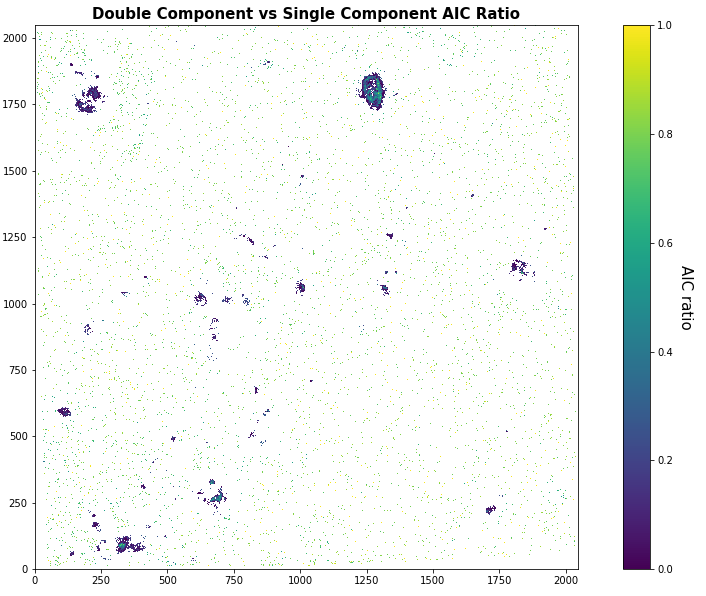}
    \caption{Left: Proxy signal-to-noise ratio versus mean absolute broadening residual (\kms) for synthetic data created to simulate a range of SNR values. For each SNR bin, we excluded outliers before calculating the mean absolute residual and standard deviation (grey y-axis error bars). Each SNR bin has a width of 1. Right: Ratio of double vs single component AIC parameters for the masked region of interested.}
    \label{fig:snr-testing}
\end{figure}
Figure \ref{fig:snr-testing} demonstrates that the network performs well for high SNR values. Thus the network is not biased for high SNR regions.
Note that the SNR value used in this section is the true signal-to-noise ratio as compared to that used in $\S$\ref{sec:M33} which is a proxy value calculated by dividing the H$\alpha$ flux by its fit uncertainty. 

In order to determine whether or not the regions of high SNR in the South West field of M33 have single or double emission components, we turn to the standard \texttt{ORCS} fitting procedure. 
We chose a small region (2x2 pixels) in a high SNR region that also has a large broadening residual (\textit{01:32:16.03, +30:48:00.71}). 
We selected pixels which fit the following prescription: have a broadening residual higher than 10 kms$^{-1}$ and a signal-to-noise ratio over 12.
We fit the H$\alpha$ and NII doublet assuming a single emission component and a double emission component. The double emission fit resulted in a statistically significantly better fit statistic. This is a strong indication that the region is best described by a double emission component rather than a single emission component. Moreover, we computed the AIC parameter for each region defined by $AIC=2n-ln(L)$, where $n$ is the number of fit parameters and $L$ is the Gaussian likelihood function (e.g. \citealt{akaike_factor_1987}; \citealt{liddle_information_2007}; \citealt{kieseppa_akaike_1997}). In our case, the likelihood is Gaussian, therefore the log-likelihood function reduces to the usual half $\chi$-squared. The right hand-size of figure \ref{fig:snr-testing} shows the ratio of the double component AIC parameter vs the single component AIC parameter defined as $exp(-(AIC_1-AIC_0)/2)$. Since the ratio is consistently below one, the double component model is favored over the single component model. 
We thus conclude that, at least in these regions, the rise in the residual value is due to the existence of double component emission. Therefore, we believe that figure \ref{fig:snr_broad} does not reflect a failure of the network in high SNR regions, but rather a failure of the network in regions with double emission components that serendipitously appear in regions of high SNR in the South West field of M33. Future work will explore the applicability of a modified network to estimate the broadening and velocity parameter in such regions.

\bibliography{SitelleML}{}
\bibliographystyle{aasjournal}



\end{document}